\documentclass[aps,twocolumn,showpacs,pre,amsmath,amssymb]{revtex4}
\usepackage{graphicx} 

\usepackage{amsfonts}

\usepackage[latin1]{inputenc}
\usepackage[T1]{fontenc}
\usepackage{xcolor,ulem} 

\usepackage{verbatim}

\begin{document}

\title{A closer look at coupled logistic maps at the edge of chaos}

\author{Ugur Tirnakli}
\email{ugur.tirnakli@ege.edu.tr}
\affiliation{Department of Physics, Faculty of Science, Ege University, 35100 Izmir, Turkey}

\author{Constantino Tsallis}
\email{tsallis@cbpf.br}
\affiliation{Centro Brasileiro de Pesquisas F\'\i sicas and National Institute for Science and 
Technology of Complex Systems, Rua Dr. Xavier Sigaud 150, 22290-180 Rio de Janeiro, RJ, Brazil\\
and\\
Santa Fe Institute, 1399 Hyde Park Road, Santa Fe, NM 87501, USA}

\date{\today}

\begin{abstract}
We focus on a linear chain of $N$ first-neighbor-coupled logistic maps at their edge of chaos in 
the presence of a common noise. This model, characterised by the coupling strength $\epsilon$ and 
the noise width $\sigma_{max}$, was recently introduced by Pluchino et al 
[Phys. Rev. E {\bf 87}, 022910 (2013)]. They detected, for the time averaged returns with characteristic 
return time $\tau$, possible connections with $q$-Gaussians, the distributions which optimise, 
under appropriate constraints, the nonadditive entropy $S_q$, basis of nonextensive statistics mechanics. 
We have here a closer look on this model, and numerically obtain probability distributions which exhibit 
a slight asymmetry for some parameter values, in variance with simple $q$-Gaussians. 
Nevertheless, along many decades, the fitting with $q$-Gaussians turns out to be numerically very 
satisfactory for wide regions of the parameter values, and we illustrate how 
the index $q$ evolves with $(N, \tau, \epsilon, \sigma_{max})$.  
It is nevertheless instructive on how careful one must be in such numerical analysis. 
The overall work shows that physical and/or biological systems that are correctly mimicked by 
the Pluchino et al model are thermostatistically related to nonextensive statistical mechanics 
when time-averaged relevant quantities are studied.
\end{abstract}


\maketitle


\section{Introduction}

Synchronization has long been observed in complex systems and extensively studied 
in the literature \cite{synch1,synch2}. In these studies, coupled maps are considered 
as an important theoretical model for these systems \cite{kaneko}.
Moreover, since many biological systems evolve in noisy environments, 
it is important to analyse  the effect of noise in such coupled maps. 
Recently, the effect of weak noise on globally coupled chaotic units has been studied 
for several systems \cite{kaneko2,kuramoto}.

The model that we focus on here is a linear chain of $N$ coupled logistic maps, 
with periodic boundary conditions, and can be written as  

\begin{equation}
x_{t+1}^i = (1-\epsilon) f\left(x_t^i\right) + 
\frac{\epsilon}{2} \left[f\left(x_t^{i-1}\right) + f\left(x_t^{i+1}\right)\right] + 
\sigma(t)
\label{overall}
\end{equation}
where $\epsilon\in [0,1]$ is the local coupling, and $\sigma(t)$ is an additive 
random noise uniformly distributed in $[0,\sigma_{max}]$, which fluctuates in time 
but is equal for all maps. Here, the $i$th logistic map at any time $t$ is given as  

\begin{equation}
f\left(x_t^{i}\right) = 1-\mu \left|x_t^i\right|^2
\end{equation}
where $\mu \in [0,2]$ is the map parameter and this function is taken in module 1 with 
sign in order to fold the iterates of the maps back into the map interval $[-1,1]$ 
if the noise takes them out of this interval. 
All elements of the system will be kept at the chaos threshold by fixing the
$\mu$ parameter at $\mu_c=1.40115518909...\,$.

Very recently, the probability distribution functions (PDFs) of the returns for this model 
have been investigated numerically by Pluchino et al.\cite{Alex2013} and fat-tailed distributions 
have been reported, which 
can be fitted by $q$-Gaussians. The physical quantity under investigation, 
the returns, is a commonly used one in areas such as turbulence \cite{turbulence}, 
finance \cite{finance,finance2}, DNA sequences \cite{DNA},
and earthquake dynamics \cite{caruso,bakar1,bakar2,ahmet} in the literature. It is defined as

\begin{equation}
\Delta d_t = d_{t+\tau} - d_t
\end{equation}
where 

\begin{equation}
d_t = \frac{1}{N} \sum_{i=1}^{N} \left|x_t^i - \left<x_t^i\right>\right| \,\, .
\end{equation}

\begin{figure}[ht]
\begin{center}
\includegraphics[width=0.97\columnwidth,keepaspectratio,clip=]{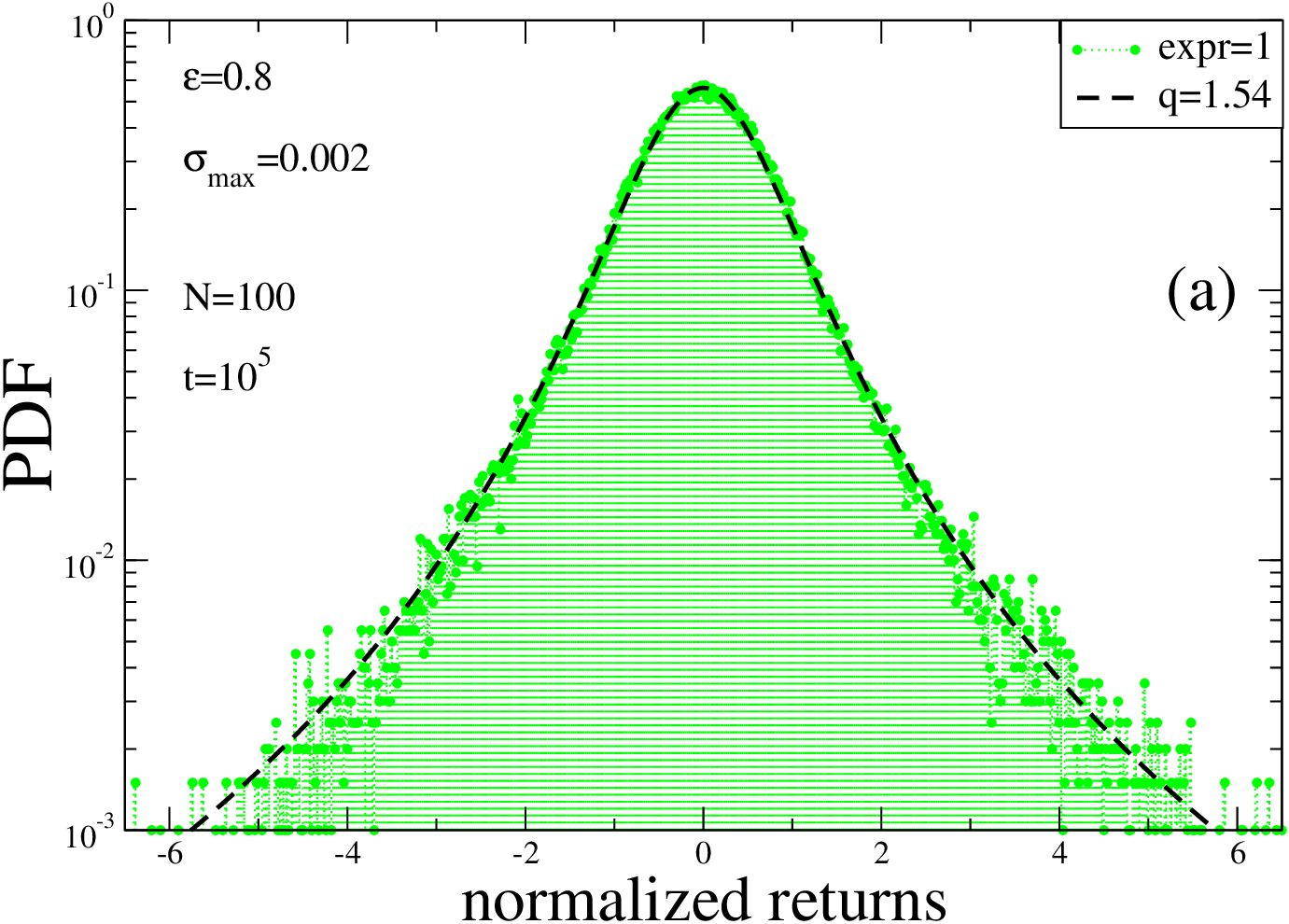} 
\includegraphics[width=0.97\columnwidth,keepaspectratio,clip=]{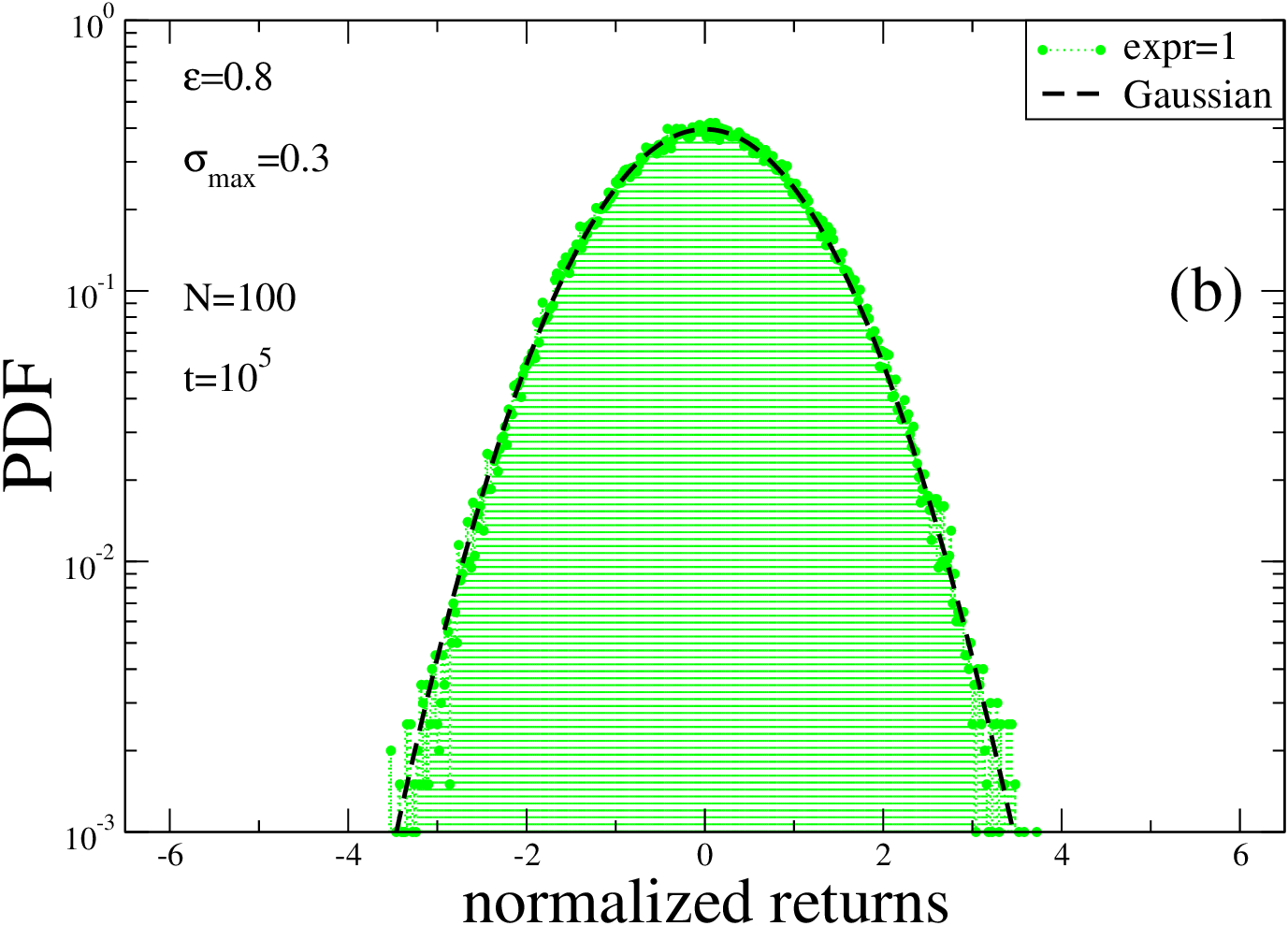} 
\end{center}
\caption{Probability distribution functions of the normalized returns for 
$N=100$ logistic maps at the edge of chaos with $\tau=32$. 
The parameter values used here are exactly those of Fig.~3a 
(case (a)) and of Fig.~3d (case (b)) 
of \cite{Alex2013}. By expr $=1$ we mean that only one experiment has been done in each of these illustrations.}
\label{fig:fig1}
\end{figure}

\begin{figure}[ht]
\begin{center}
\includegraphics[width=0.97\columnwidth,keepaspectratio,clip=]{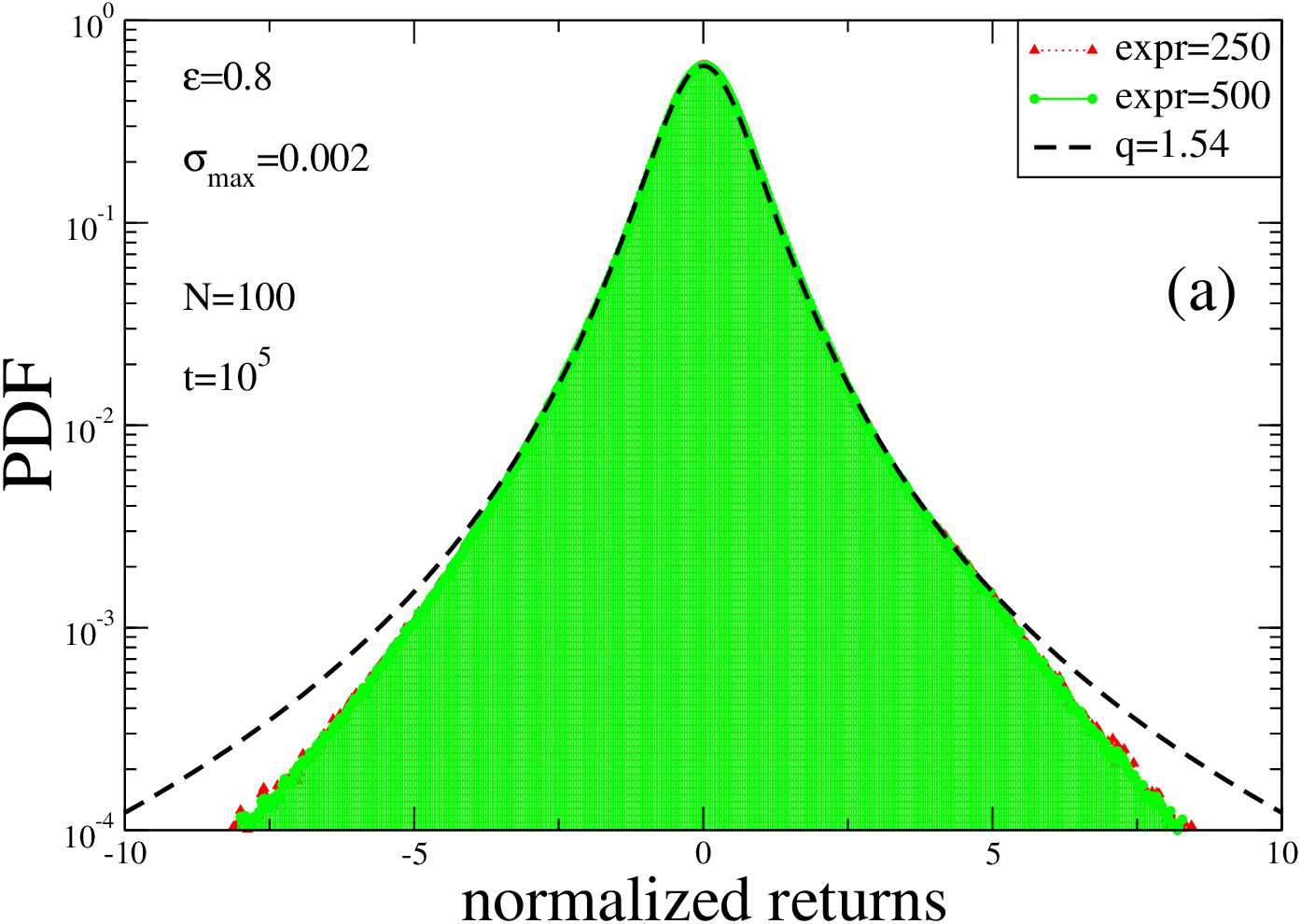}
\includegraphics[width=0.97\columnwidth,keepaspectratio,clip=]{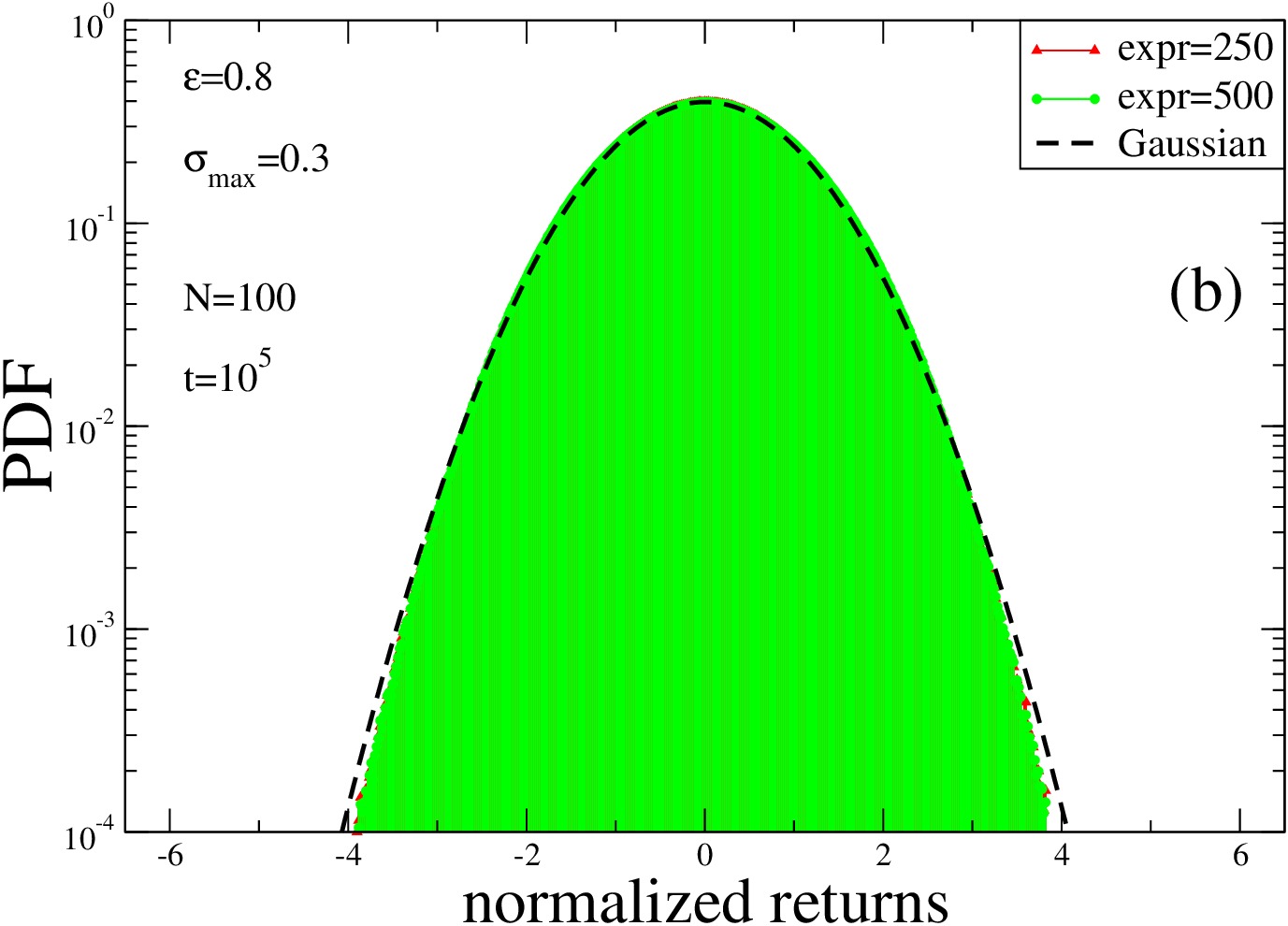} 
\end{center}
\caption{Probability distribution functions of the normalized returns for 
$N=100$ logistic maps at the edge of chaos with  $\tau=32$ for $\sigma_{max}=0.002$ (a) 
and $\sigma_{max}=0.3$ (b). In order to analyse the tails 
of the distribution, large number of experiments is used. The size of the 
time series therefore equals to the number of experiments times the time 
steps used in each experiment (i.e., expr$\times 10^5$).}
\label{fig:fig2}
\end{figure}

The main result of \cite{Alex2013} has been given in Fig.~3 of that paper where the 
PDFs of the normalized returns (normalized to the standard deviation of the overall 
sequence) have been plotted for various $\sigma_{max}$ values. We have reproduced 
here in our Fig.~\ref{fig:fig1}  two representative cases, namely, $\sigma_{max}=0.002$ 
and $\sigma_{max}=0.3$ to be compared with Fig.~3a and Fig.~3d of \cite{Alex2013}, 
respectively. It is evident that the $q$-Gaussian curves with predicted $q$ values are 
very reasonable for these examples ($q=1.54$ for $\sigma_{max}=0.002$ and $q=1$ for $\sigma_{max}=0.3$). 
However, the distributions span less than 3 decades since 
the time series that have been used are relatively short ($t=10^5$). Therefore it is interesting to 
check the behaviour 
with longer time series in order to span more decades of the PDF. This is what we have done: see
Fig.~\ref{fig:fig2}, which has been obtained by considering a large number of experiments (250 and 
500). This means that the time series analysed are as long as $250\times 10^5$ and 
$500\times 10^5$ respectively. As seen clearly from Fig.~\ref{fig:fig2}, when one more 
decade is exhibited for the tails, the departure from the {\it a priori} admissible $q$-Gaussians becomes 
clearly visible.

\begin{figure}[ht]
\begin{center}
\includegraphics[width=0.97\columnwidth,keepaspectratio,clip=]{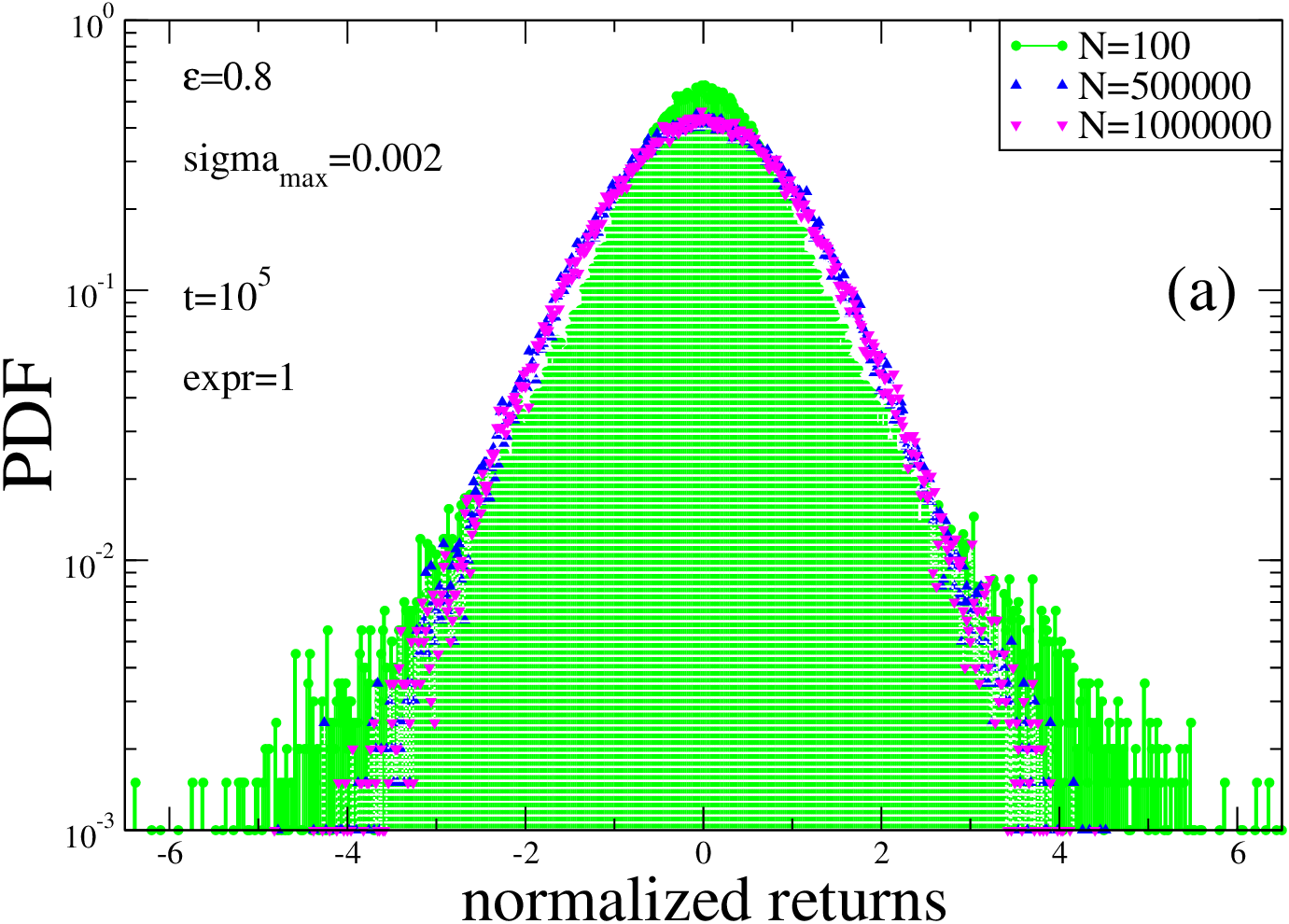}
\includegraphics[width=0.97\columnwidth,keepaspectratio,clip=]{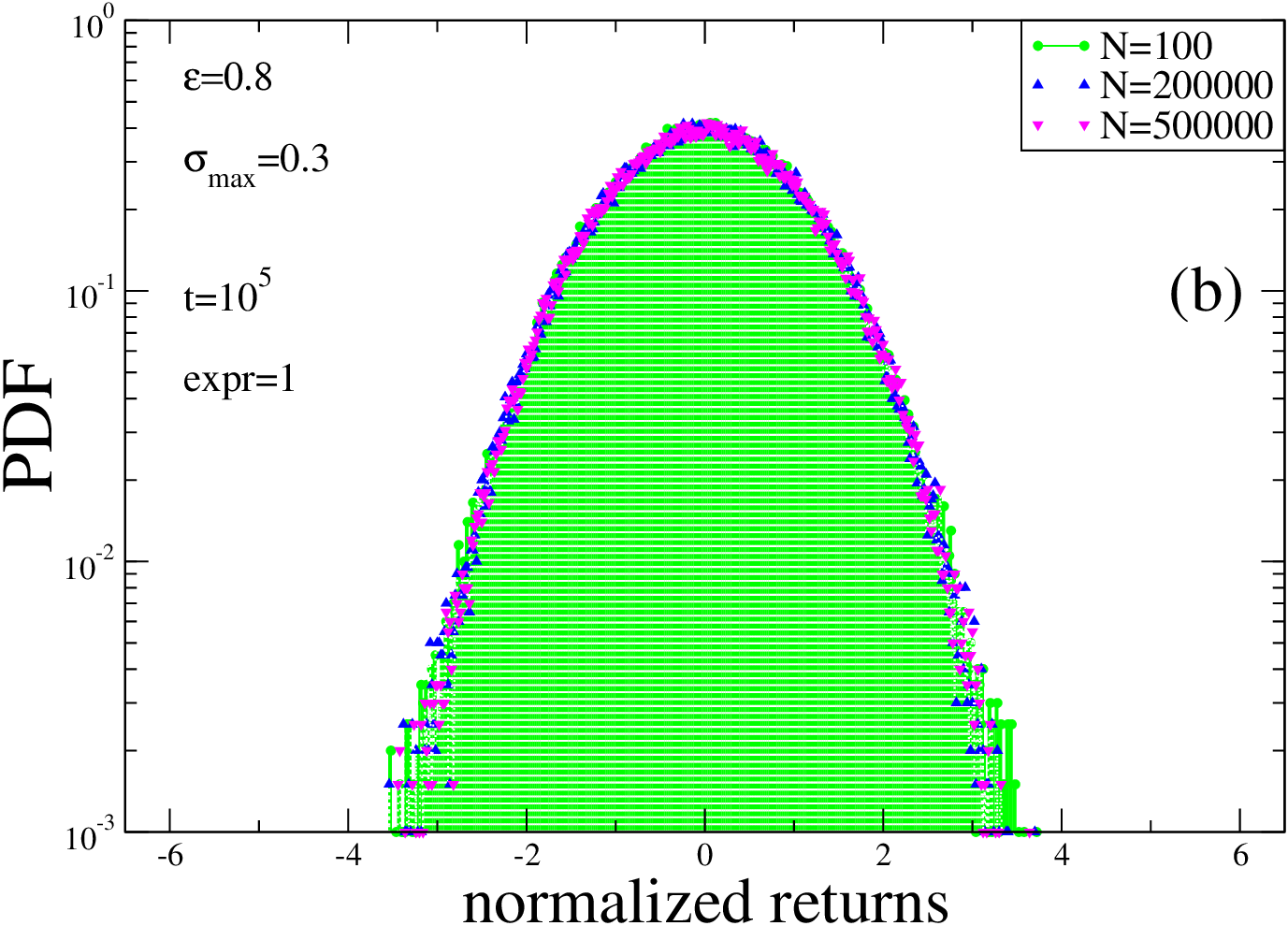} 
\end{center}
\caption{Probability distribution functions of the normalized returns for 
$N$ logistic maps at the edge of chaos with  $\tau=32$ for $N=100$, $N=100000$ and $N=500000$. 
It is evident that, as $N$ increases, the asymptotic PDFs are attained 
slowly for small $\sigma_{max}$ values (i.e., case (a)), whereas it is quite quick 
for larger $\sigma_{max}$ values (i.e., case (b)).}
\label{fig:fig3}
\end{figure}

In addition to this, it is also interesting to check whether these typical numerical PDFs are asymptotic 
in the sense of the $N\to\infty$ limit.
If this is the case, one would expect the PDF 
curves to remain basically invariant as the size $N$ of the system increases. We have checked it: 
see two illustrative cases in Fig.~\ref{fig:fig3}. 
It is easily seen that the $N=100$ size cannot be considered as sufficient for the analysis 
of the asymptotic behaviour, especially for small values of $\sigma_{max}$.

\begin{figure*}[ht]
\begin{center}
\includegraphics[scale=0.35]{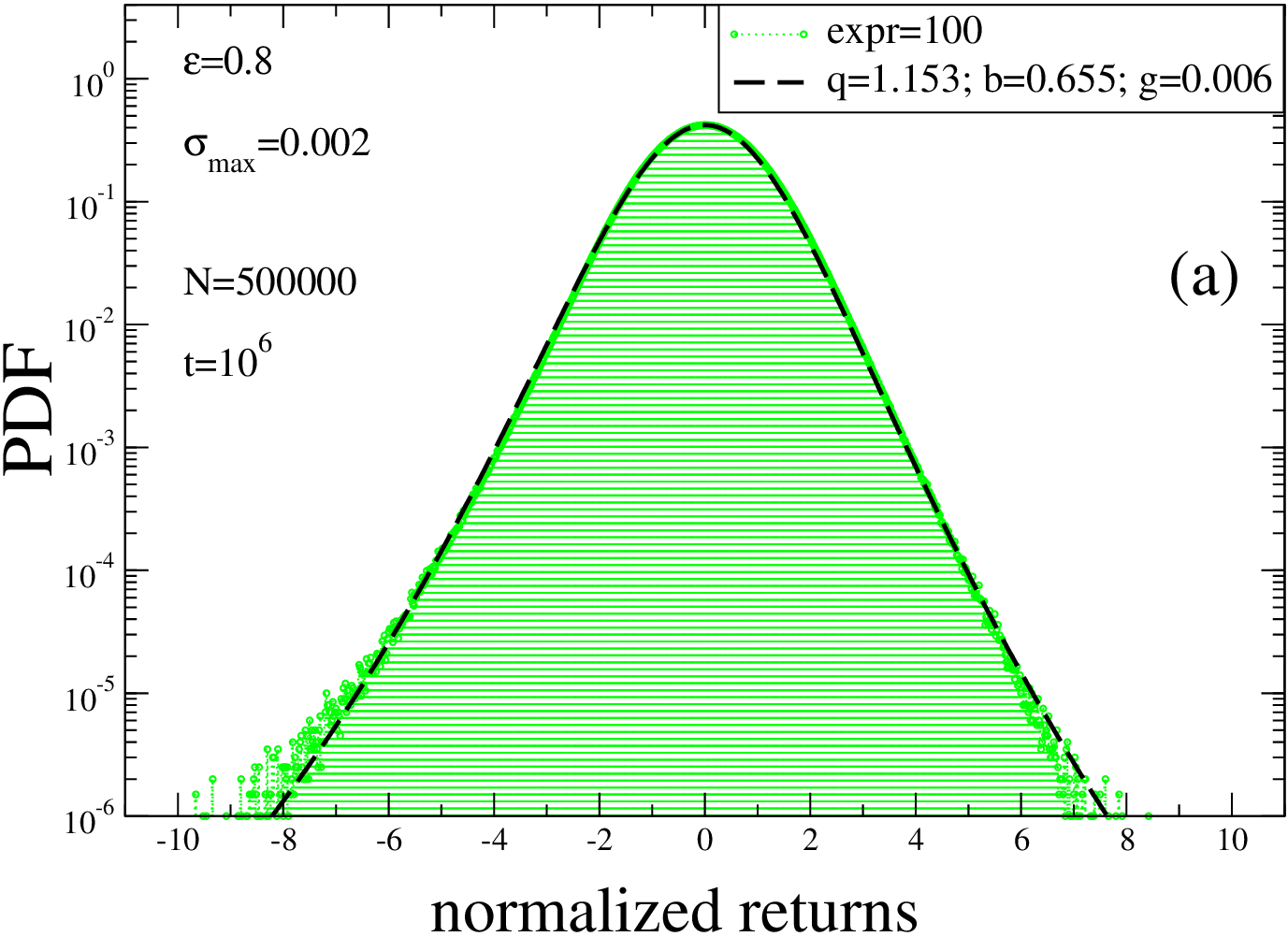}
 \hspace{0.3cm}
\includegraphics[scale=0.35]{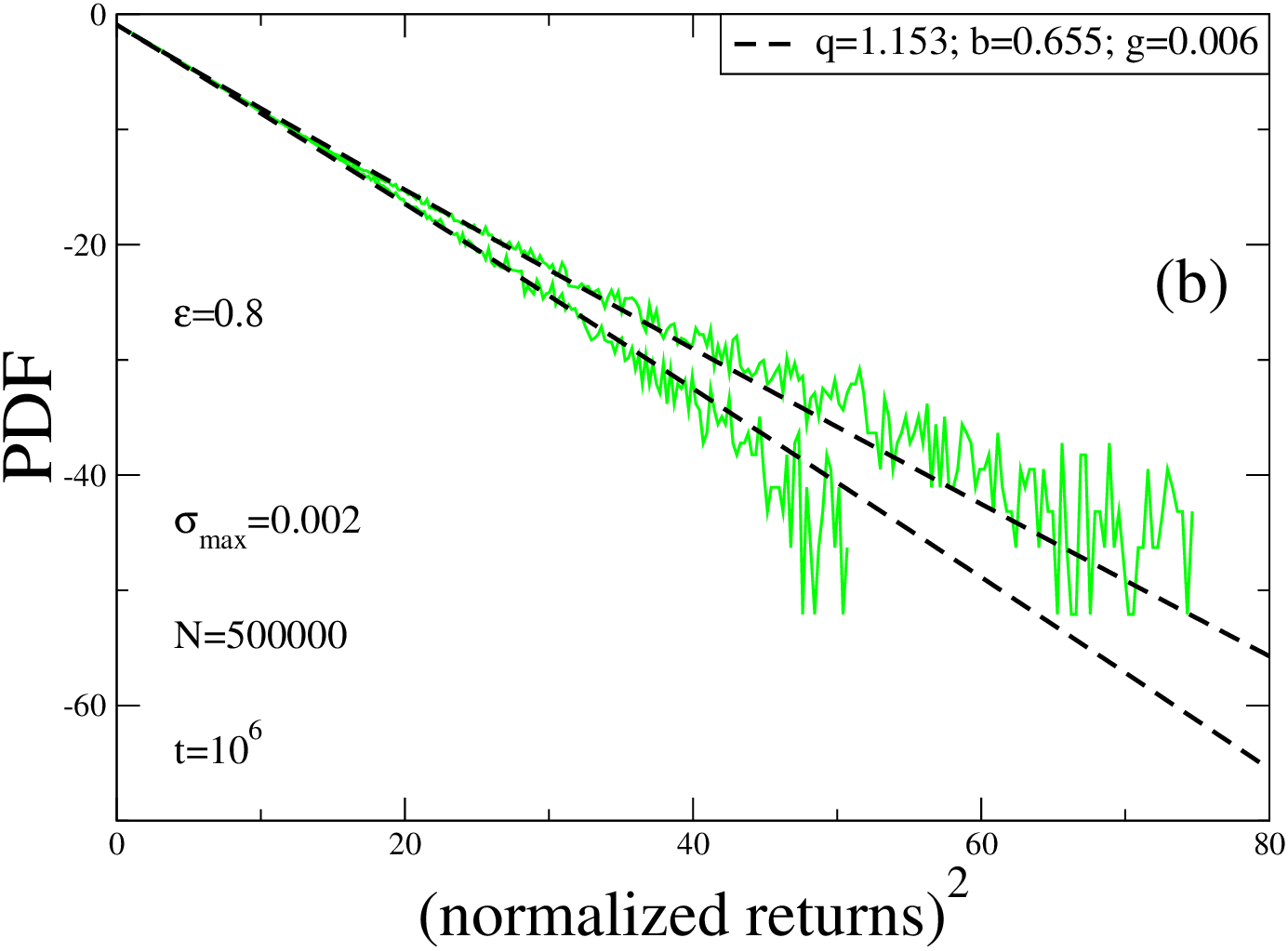} 
\\[0.5cm]
\includegraphics[scale=0.35]{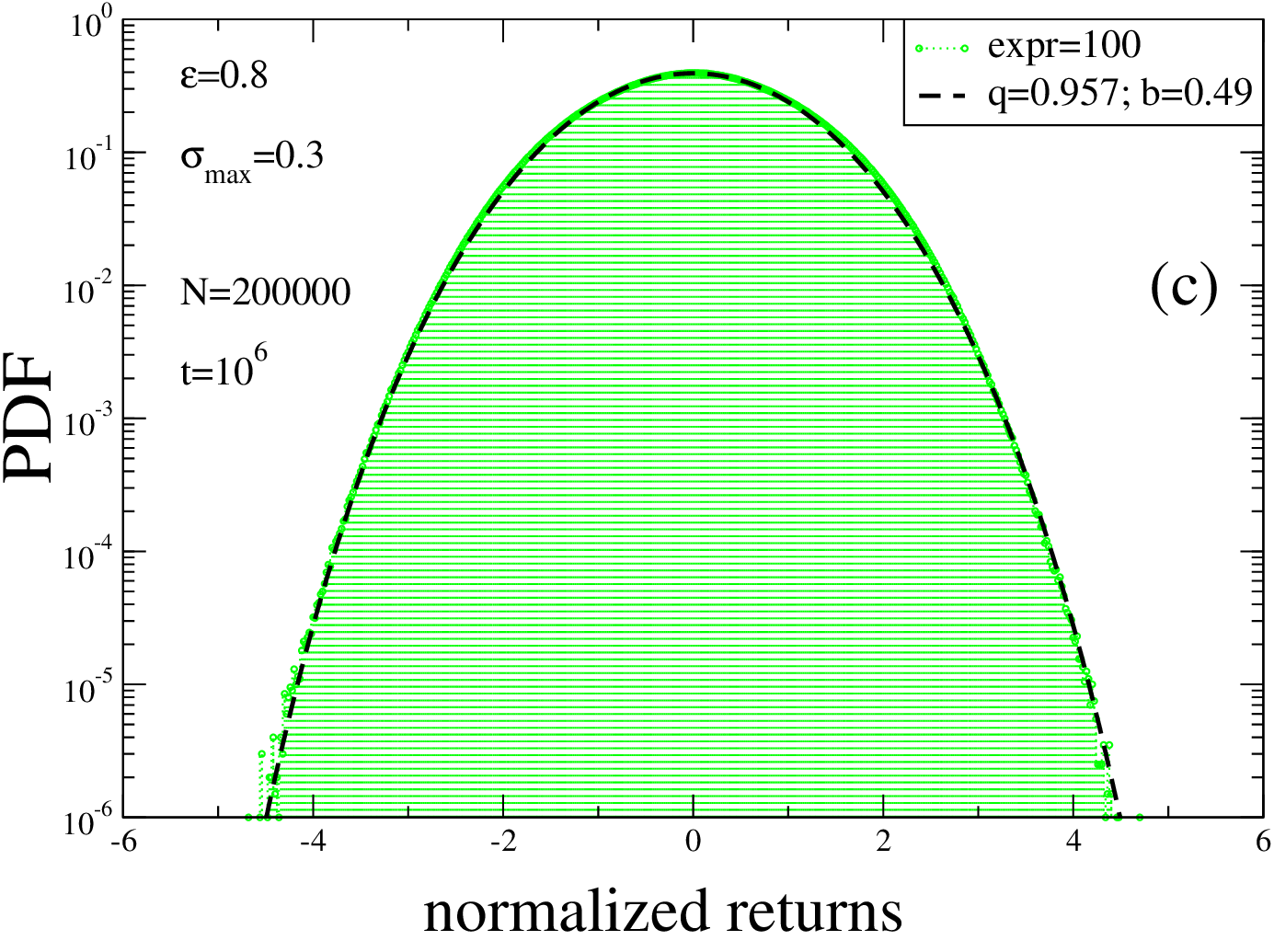}
\hspace{0.3cm}
\includegraphics[scale=0.35]{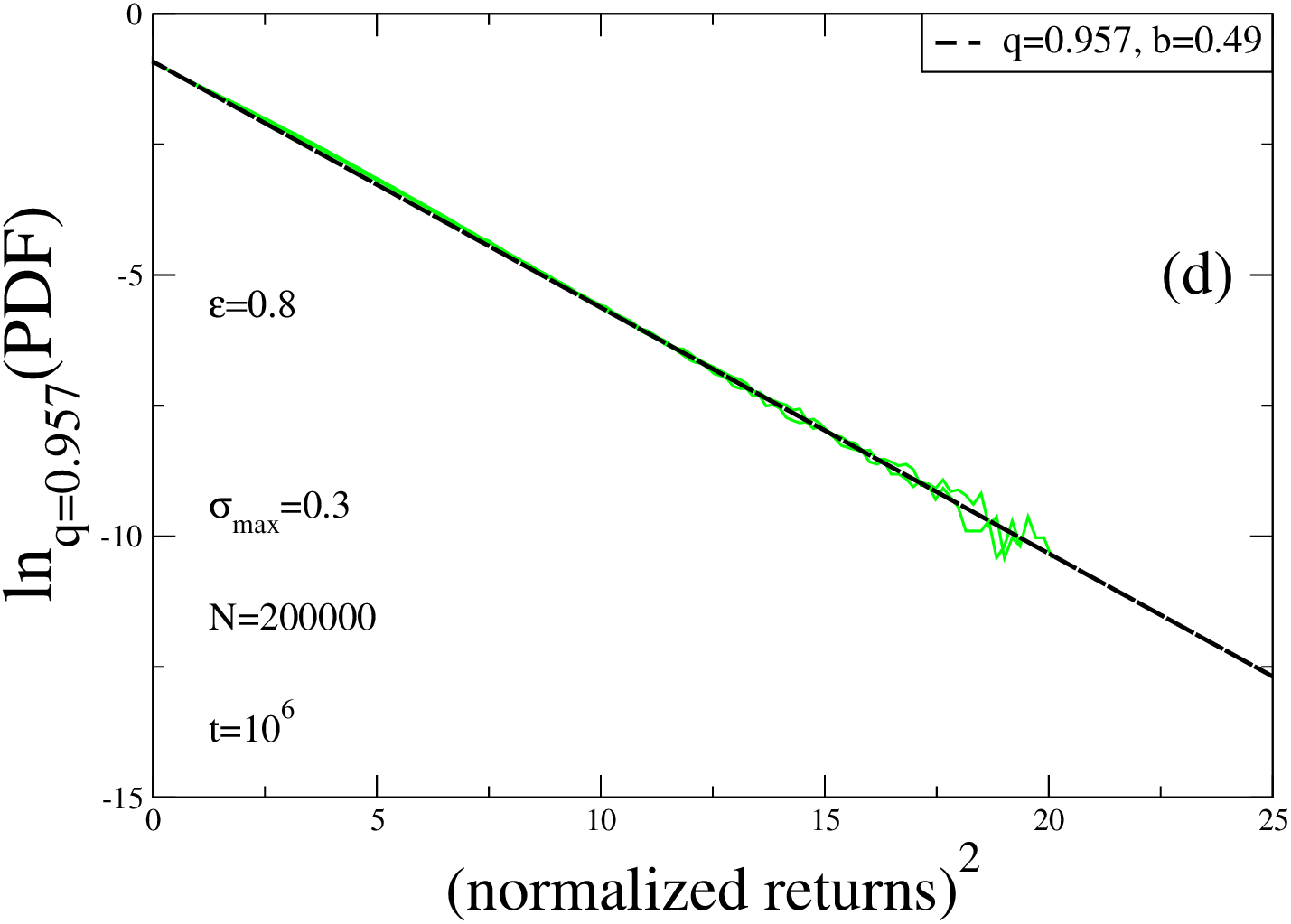} 
\end{center}

\caption{Asymptotic PDFs of the normalized returns with large $N$ and large 
number of time steps for $\sigma_{max}=0.002$ (a) and $\sigma_{max}=0.3$ (c). 
For both cases $\tau=32$. The $q$-logarithms versus the square of normalized returns 
have already been given for both cases in (b) and (d) respectively.}
\label{fig:fig4}
\end{figure*}

It is therefore clear at this stage that, if we are particularly interested in studying the 
asymptotic behaviour of PDFs of the normalized returns of this system, both
the size $N$ of the system and the length of the time series must be quite large. To further refine 
numerically the present analysis, we have also studied the same two cases with large enough value for 
$N$ ($N=500000$ for $\sigma_{max}=0.002$. and $N=200000$ for $\sigma_{max}=0.3$), and with a time series of 
$100\times 10^6$ steps. These conditions are  sufficient to observe the behaviour up to almost 6 decades. 
The results are given in Fig.~\ref{fig:fig4}. The strong departure from the $q$-Gaussians 
with early predicted $q$ values becomes now very obvious. Moreover, a slight but neat asymmetry emerges 
for small $\sigma_{max}$. In order to fit this kind of behavior, one can use a non-symmetric extension 
of $q$-Gaussian, namely given by 

\begin{equation}
p \propto \left[1-(1-q)(bx^2+gx^3)\right]^{1/(1-q)} \,,
\end{equation}
or even
\begin{equation}
p \propto \left[1-(1-q)(bx^2+gx^3+hx^4)\right]^{1/(1-q)} \,.
\end{equation}
These expressions recover the symmetric $q$-Gaussian for $g=h=0$. Strictly speaking, the case $g \ne 0$ 
and $h=0$ is mathematically inadmissible since it would lead to a runaway to infinity. In contrast, 
the case $g \ne 0$ and $h>0$ is perfectly admissible. However, in practice, unless we explore very large 
values of $x$, we can consider $h=0$ with no sensible damage.
A very good fitting has been obtained by using $q=1.153$, $b=0.655$, $g=0.006$ (with $h=0$) as illustrated 
in Fig.~\ref{fig:fig4}a. 
The representation of the same data using the $q$-logarithmic function, defined as 
$\ln_q y \equiv (y^{1-q}-1)/(1-q) \,\, (y>0; \, \ln_1y=\ln y)$, has also been 
given in Fig.~\ref{fig:fig4}b. 
We notice that, in the asymptotic regime, the PDF for large values $\sigma_{max}$ approaches 
a $q$-Gaussian with $q<1$, rather than a Gaussian, as first advanced  in Ref.\cite{Alex2013} far from 
the asymptotic regime: see Figs.~\ref{fig:fig4}c and \ref{fig:fig4}d.

\section{$q$-Gaussian approximants}

Let us now attempt a wider look at the same problem, by extending the 
preliminary results given in \cite{Alex2013}. A simple $q$-Gaussian approximant describes fairly well 
the numerical data unless we focus on a parameter value where an asymmetry emerges in the PDF. 
The index $q$ is generically expected to depend on $(\epsilon, \sigma_{max}, \tau)$.

First of all, we need to define the procedure that we are employing in order to predict the 
best value of $q$ parameter of the related $q$-Gaussian at each tuple. 
The $q$-Gaussians can be obtained by optimising the entropy 
$S_q=\left(1-\sum_{i=1}^W p_i^q\right)/(q-1)$ \cite{tsallis1} and can be 
given for $-\infty <q<3$ as

\begin{equation}
p = A_q \sqrt{B_q} \, e_q^{-B_q x^2}
\label{qGauss}
\end{equation}
where $e_q^y$, known as $q$-exponential, is defined by

\begin{equation}
e_q^y =\left\{\begin{array}{ll}
                              \left[1+(1-q)y\right]^{1/(1-q)} \,\,\, ,&\,\,\, 
                              \mbox{$1+(1-q)y\geq 0$}\\
                              \,\,\,\,\,\,\,\,\,\,\,\,\,\,\,\,\,\,  0 \,\,\,\,\,\,\,\,\,\,\,\,\,\,\,\,\,\,\,\,\,\,\,\,\,\,\,\,\,\,\,\,\,,
                              &\,\,\,\,\,\,\,\,\,\,\,\,\,\,\,\,\,\, \mbox{else} \,.
                                \end{array}
                           \right.
\end{equation}
By inverting this function we obtain the $q$-logarithm function defined above.

The $q$-dependent coefficient $A_q$  in Eq. (\ref{qGauss}) is given by

\begin{equation}
A_q =\left\{\begin{array}{ll}
                              \frac{\Gamma\left(\frac{5-3q}{2-2q}\right)}{\Gamma\left(\frac{2-q}{1-q}\right)} 
                              \sqrt{\frac{1-q}{\pi}}\,\,\, ,
                              &\,\,\,\mbox{$q<1$}\\
                              \frac{1}{\sqrt\pi} \,\,\, ,&\,\,\, \mbox{$q=1$}\\
                                \frac{\Gamma\left(\frac{1}{q-1}\right)}{\Gamma\left(\frac{3-q}{2q-2}\right)} 
                              \sqrt{\frac{q-1}{\pi}}\,\,\, ,
                               &\,\,\,\mbox{$1<q<3$} \,.
                                \end{array}
                           \right.
\end{equation}
The quantity $B_q$ characterizes the PDF width $w$ of the distribution as follows:
\begin{equation}
B_q = \frac{1}{(3-q)w} \,\, ,
\end{equation}
where $w$ is, for  $q<5/3$, related to the standard deviation $s$ through 
\begin{equation}
(5-3q) s^2 = (3-q) w^2 \,\, . 
\end{equation}

\begin{figure*}[ht]
\begin{center}
\includegraphics[scale=0.35]{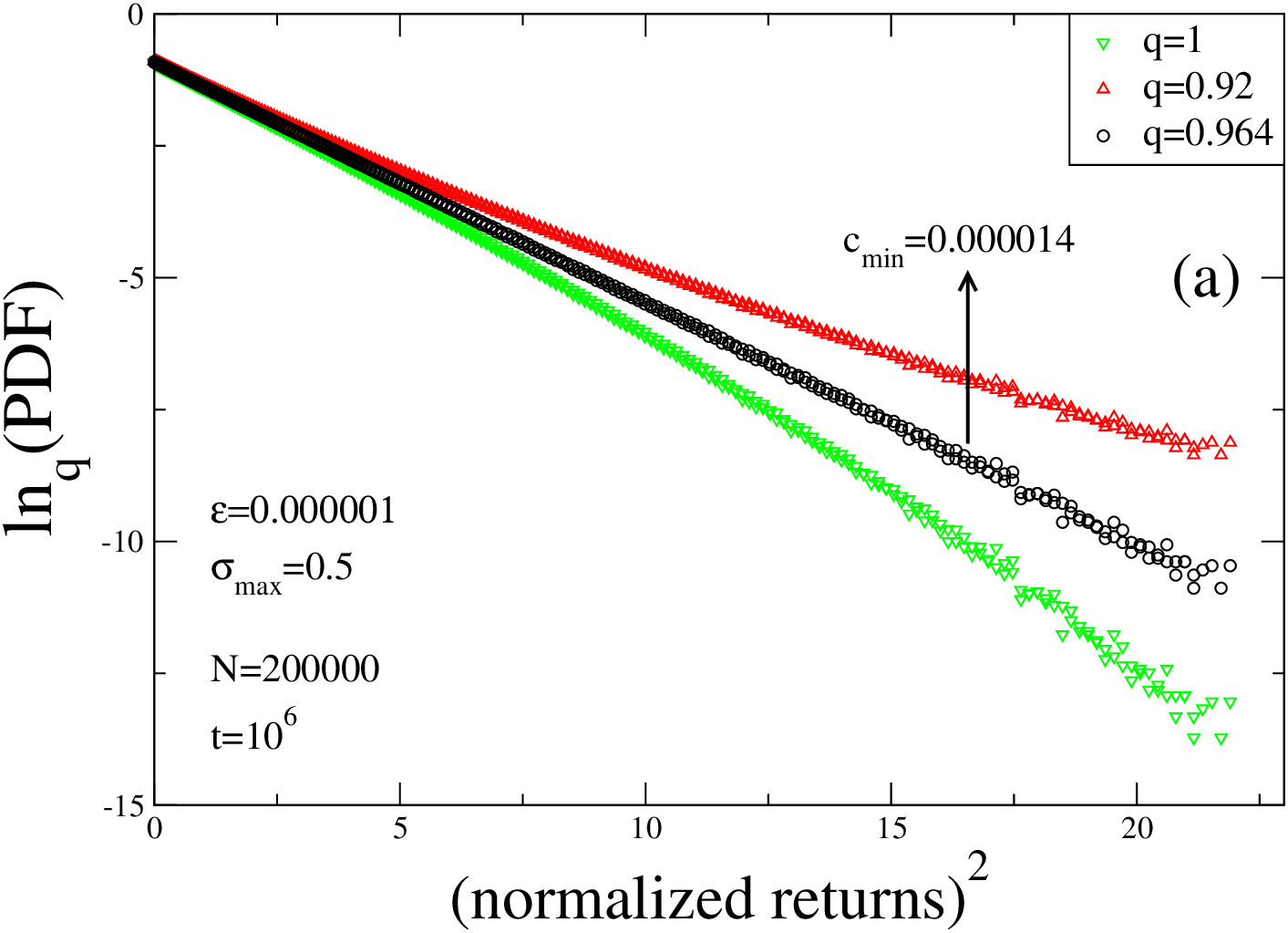}
\hspace{0.3cm}
\includegraphics[scale=0.35]{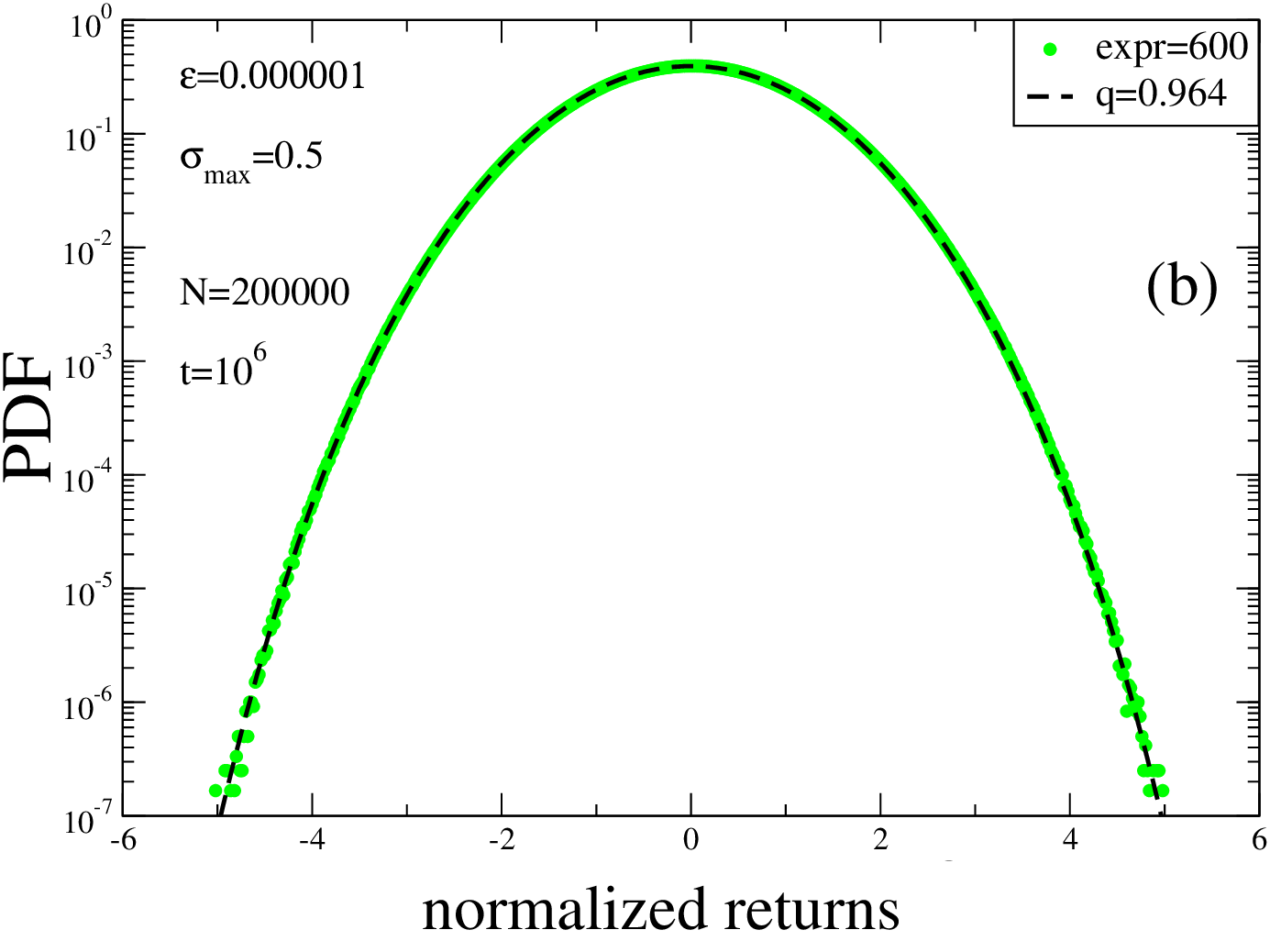} 
\\[0.5cm]
\includegraphics[scale=0.35]{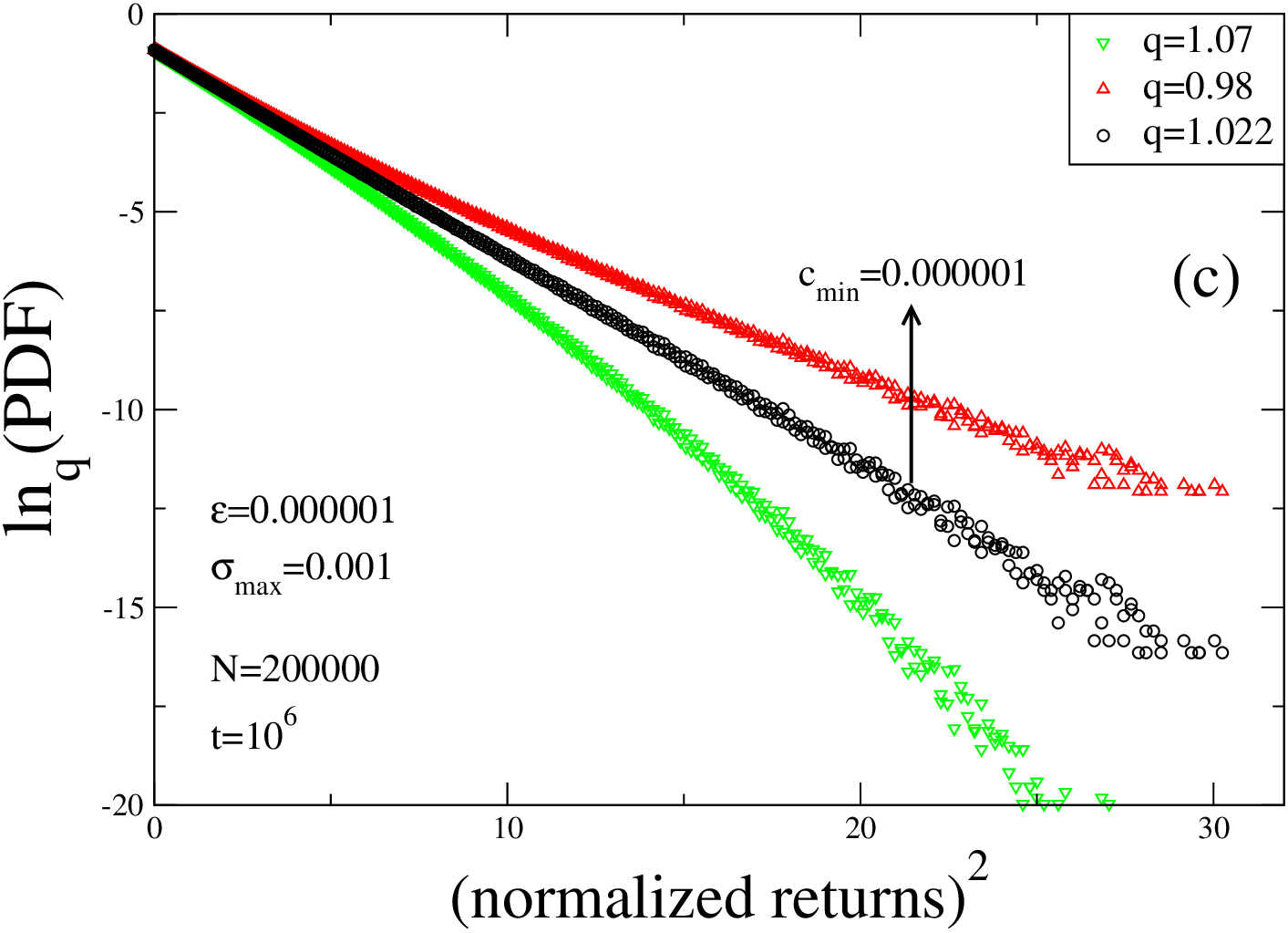}
\hspace{0.3cm}
\includegraphics[scale=0.35]{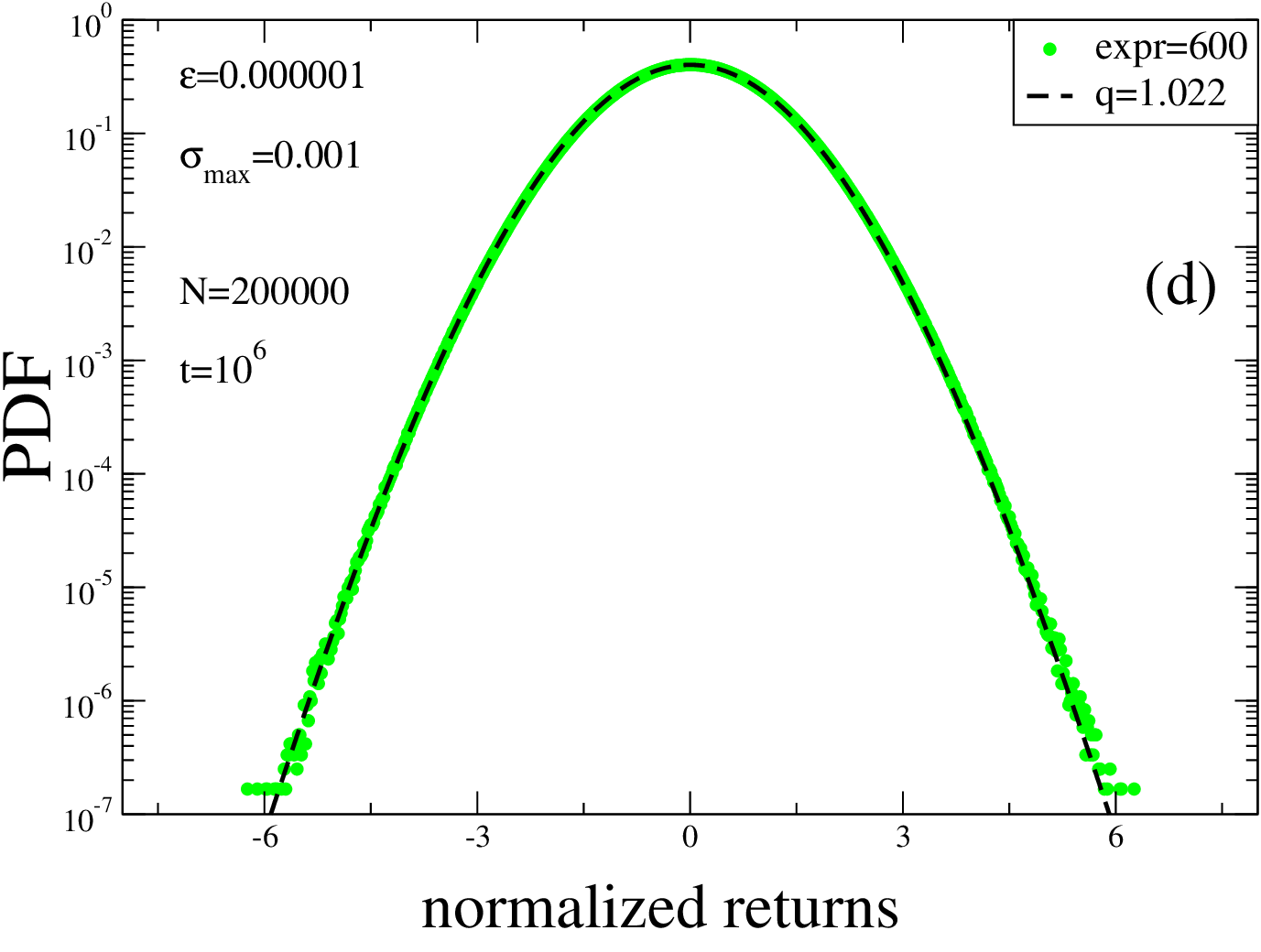} 
\end{center}
\caption{Two typical examples of choosing the most appropriate $q$ value using the 
quadratic function for a case $q<1$ (a) and for a case $q>1$ (c). 
The appropriate $q$ value is taken as the one that gives the 
smallest $\gamma$ value (denoted as $\gamma_{min}$). 
The probability density of the same examples using the $q$ value obtained 
from the explained procedure for the case $q<1$ (b) and for the case 
$q>1$ (d).}
\label{fig:fig5}
\end{figure*}

\begin{figure}
\begin{center}
\includegraphics[width=0.97\columnwidth,keepaspectratio,clip=]{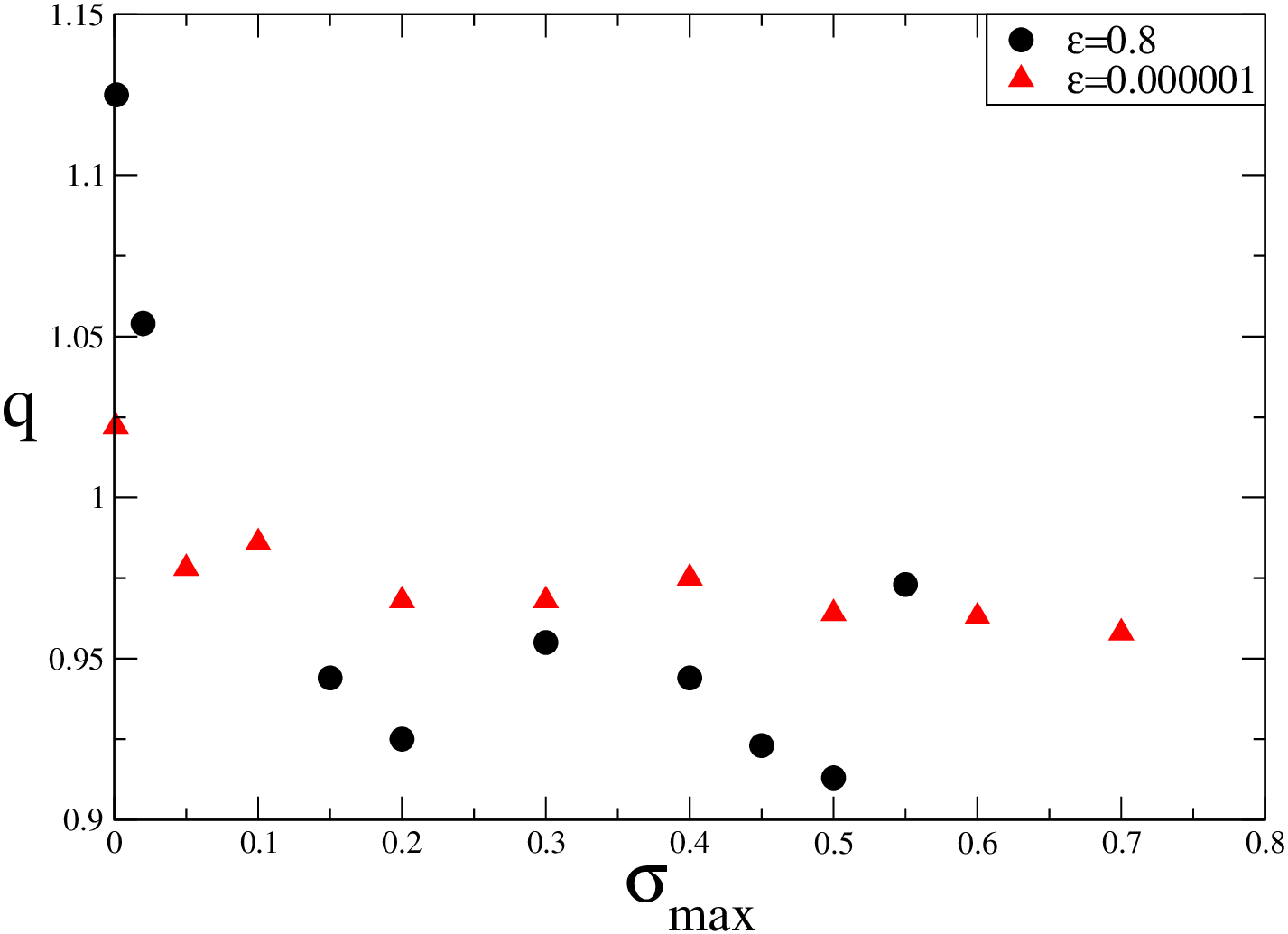}
\caption{The behaviour of the index $q$  as a function of $\sigma_{max}           
$ for two typical values 
of $\epsilon$.}
\label{fig:fig6}
\end{center}
\end{figure}

In our simulations, there are only two free parameters, namely $q$ and $B_q$, to be adjusted under 
the assumption that the number of decades that we are observing still shows a symmetric PDF.  
For a given set $(\epsilon, \sigma_{max}, \tau)$, we adjust the best value of $q$ 
by looking at the curves $\ln_q$(PDF) versus 
(normalized returns)$^2$. The procedure consists in determining the value of $q$ 
which provides the best approach to a straight line. More precisely, we fit the curves resulting 
from various values of $q$ with a quadratic function

\begin{equation}
y = \alpha + \beta x + \gamma x^2 \,\, , 
\end{equation}
and choose the value of $q$ which provides the minimal value for the parameter $\gamma$. 
Then, for this $q$ value, 
we calculate $A_q$ and $B_q$ parameters from the time series. Finally, using all these obtained 
parameter values, we plot the best $q$-Gaussian for the chosen set $(\epsilon, \sigma_{max}, \tau)$. 
Two typical examples are given in Fig.~\ref{fig:fig5}. 
For these examples, 
the minimal value of $\gamma$, denoted by $\gamma_{min}$, have been obtained for $q=0.964$ and $q=1.022$, 
respectively. Then these values of $q$ have been used to determine the values of the $A_q$ and 
$B_q$ parameters. Finally, the PDFs of the normalized returns are approached as shown in 
Fig.~\ref{fig:fig5}b and 5d.
It is evident from the figure that the best $q$-Gaussian approximants corroborate 
the behaviour of the system for more than 6 decades in each case. 
The obtained results are summarized in Table I and plotted in Fig.~\ref{fig:fig6} for
typical values of $(\epsilon,\sigma_{max})$. Unless the $\sigma_{max}$ values are very small, 
the tendency is always towards a $q$-Gaussian with $q<1$. For very small $\sigma_{max}$, 
$q>1$ are also observed but for such cases the asymmetry becomes important.

\begin{figure}
\begin{center}
\includegraphics[width=0.97\columnwidth,keepaspectratio,clip=]{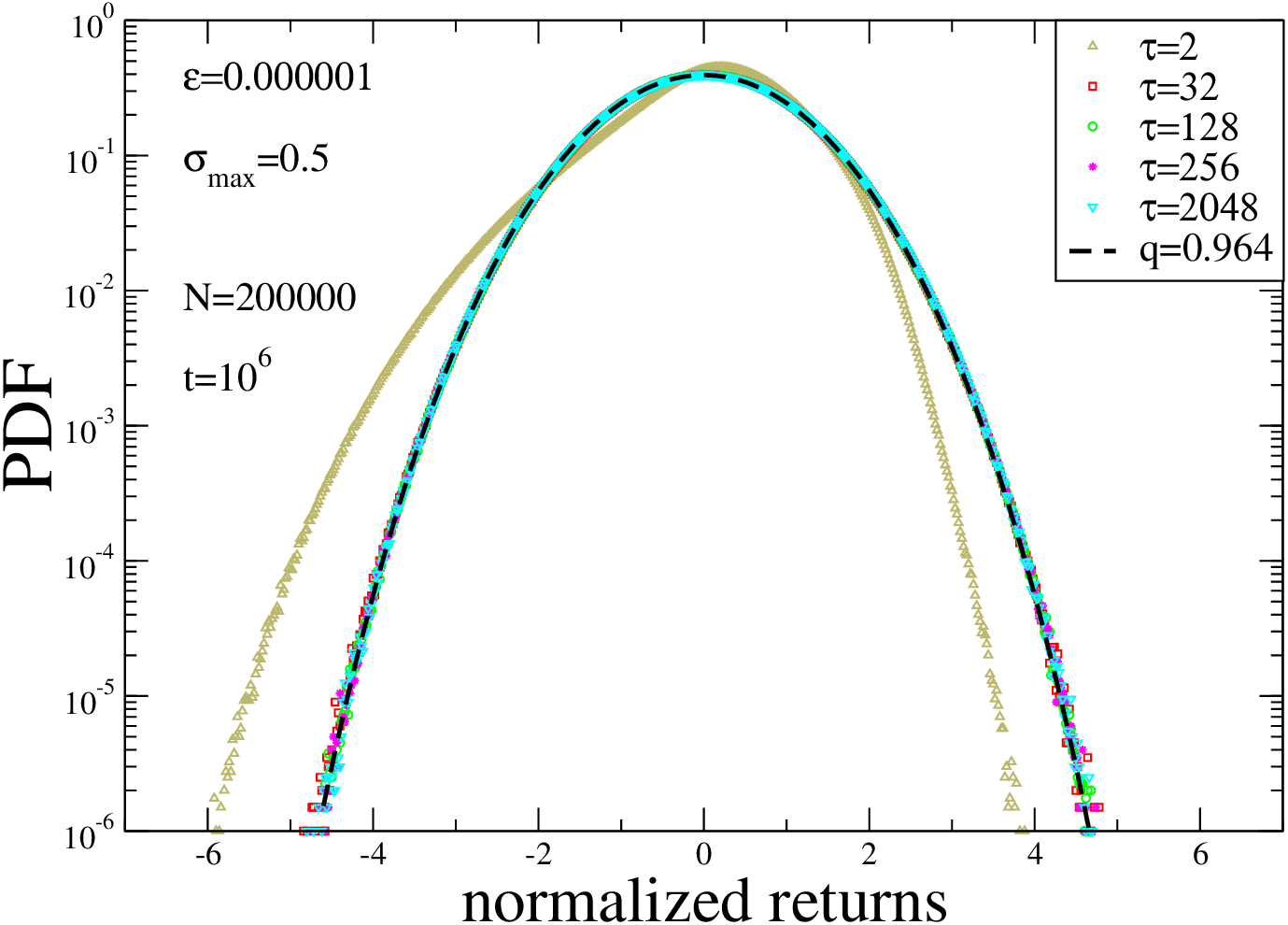}
\caption{The effect of $\tau$ on the behavior of the PDF.}
\label{fig:fig7}
\end{center}
\end{figure}

In our simulations, we prefer to set $\tau=128$ since it is evident from 
Fig.~\ref{fig:fig7} that if $\tau$ is not very small the PDF remains the same as 
$\tau$ is increased.

\begin{table}
\centering
{%
\begin{tabular}{|c|c|c|c|}
\hline 
\hline
\multicolumn{4}{|c|}{$\epsilon=0.000001$} \\ 
\hline
$\sigma_{max}$  & $A_q\sqrt{B_q}$ & $B_q$ & $q$ \\ 
\hline 
\hline 
0.001& 0.4023 & 0.5171 & 1.022 \\ 
\hline 
0.05 & 0.3957 & 0.4840 & 0.978 \\ 
\hline 
0.1  & 0.3989 & 0.4897 & 0.986 \\ 
\hline 
0.2  & 0.3944 & 0.4771 & 0.968 \\ 
\hline 
0.3  & 0.3944 & 0.4771 & 0.968 \\ 
\hline 
0.4  & 0.3953 & 0.4819 & 0.975 \\ 
\hline 
0.5  & 0.3938 & 0.4744 & 0.964 \\ 
\hline 
0.6  & 0.3937 & 0.4737 & 0.963 \\ 
\hline 
0.7  & 0.3930 & 0.4704 & 0.958 \\ 
\hline 
\end{tabular}}%
\qquad\qquad
{%
\begin{tabular}{|c|c|c|c|}
\hline 
\hline
\multicolumn{4}{|c|}{$\epsilon=0.8$} \\ 
\hline
$\sigma_{max}$  & $A_q\sqrt{B_q}$ & $B_q$ & $q$ \\ 
\hline 
\hline 
0.0015 & 0.4214 & 0.6154 & 1.125 \\ 
\hline 
0.02   & 0.4077 & 0.5441 & 1.054 \\ 
\hline 
0.15   & 0.3912 & 0.4613 & 0.944 \\ 
\hline 
0.2    & 0.3888 & 0.4494 & 0.925 \\ 
\hline 
0.3    & 0.3926 & 0.4684 & 0.955 \\ 
\hline 
0.4    & 0.3912 & 0.4613 & 0.944 \\ 
\hline 
0.45   & 0.3885 & 0.4482 & 0.923 \\ 
\hline 
0.5    & 0.3873 & 0.4423 & 0.913 \\ 
\hline 
0.55   & 0.3950 & 0.4805 & 0.973 \\ 
\hline 
\end{tabular}}

\caption{For $\tau=128$ and two typical values of $\epsilon$, the values of the parameters 
$A_q\sqrt{B_q}$, $B_q$ and $q$ as $\sigma_{max}$ varies are indicated.}
\end{table}

We have also checked the effect of $\epsilon$ for a fixed value of {$\sigma_{max}$. 
A clear tendency of increasing $q$ values as $\epsilon$ is decreasing has been observed 
with a saturation value around $q\simeq 0.965$ as given in Table II and plotted in Fig.~\ref{fig:fig8}.

\begin{table}
\begin{tabular}{|c|c|c|c|}
\hline 
\hline 
\multicolumn{4}{|c|}{$\sigma_{max}=0.5$} \\ 
\hline
$\epsilon$  & $A_q\sqrt{B_q}$ & $B_q$ & $q$ \\ 
\hline 
\hline 
0.000001 & 0.3938 & 0.4744 & 0.964 \\ 
\hline 
0.00001  & 0.3937 & 0.4737 & 0.963 \\ 
\hline 
0.0001   & 0.3941 & 0.4757 & 0.966 \\ 
\hline 
0.001    & 0.3939 & 0.4751 & 0.965 \\ 
\hline 
0.01     & 0.3939 & 0.4751 & 0.965 \\ 
\hline 
0.1      & 0.3914 & 0.4625 & 0.946 \\ 
\hline 
0.2      & 0.3895 & 0.4531 & 0.931 \\ 
\hline 
0.5      & 0.3878 & 0.4446 & 0.917 \\ 
\hline 
0.8      & 0.3873 & 0.4423 & 0.913 \\ 
\hline 
\end{tabular} 
\caption{For $\tau=128$ and a typical value of $\sigma_{max}$, the values of the parameters 
$A_q\sqrt{B_q}$, $B_q$ and $q$ as $\epsilon$ varies are indicated.}
\end{table}

\begin{figure}
\begin{center}
\includegraphics[width=0.97\columnwidth,keepaspectratio,clip=]{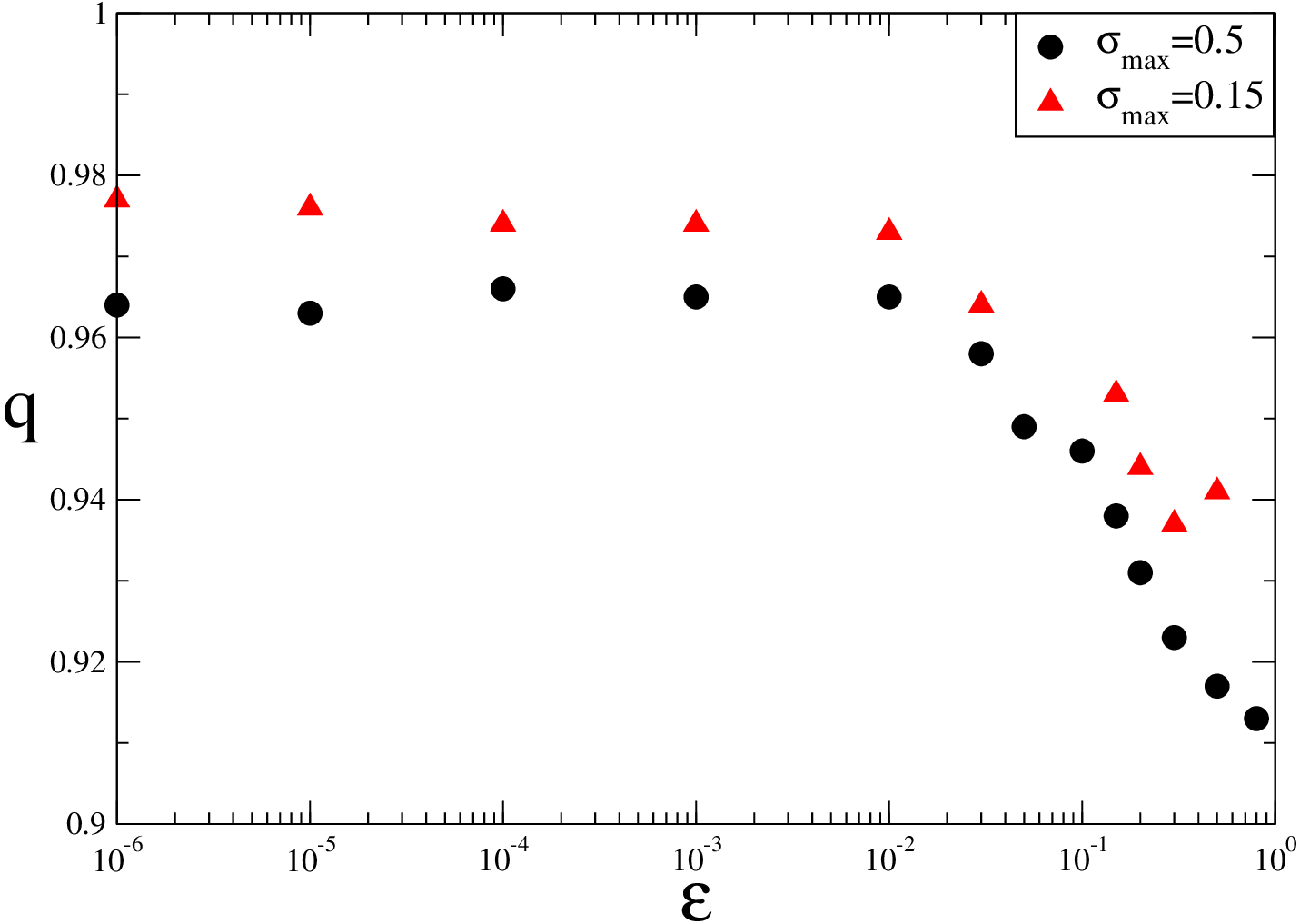}
\caption{The behaviour of the index $q$ as a function of $\epsilon$ for a typical value of
$\sigma_{max}$.}
\label{fig:fig8}
\end{center}
\end{figure}

Finally, we have simulated the case $\mu=2$ without noise. As expected, in this case, a clear Gaussian 
with 6 decades can be seen in Fig.~\ref{fig:fig9} since all individual logistic maps are at their 
mostly chaotic points.  
Whenever a nonzero noise term is included, the system is not able to achieve the Gaussian due to 
the persistence of the noise on each map.

\begin{figure}
\begin{center}
\includegraphics[width=0.97\columnwidth,keepaspectratio,clip=]{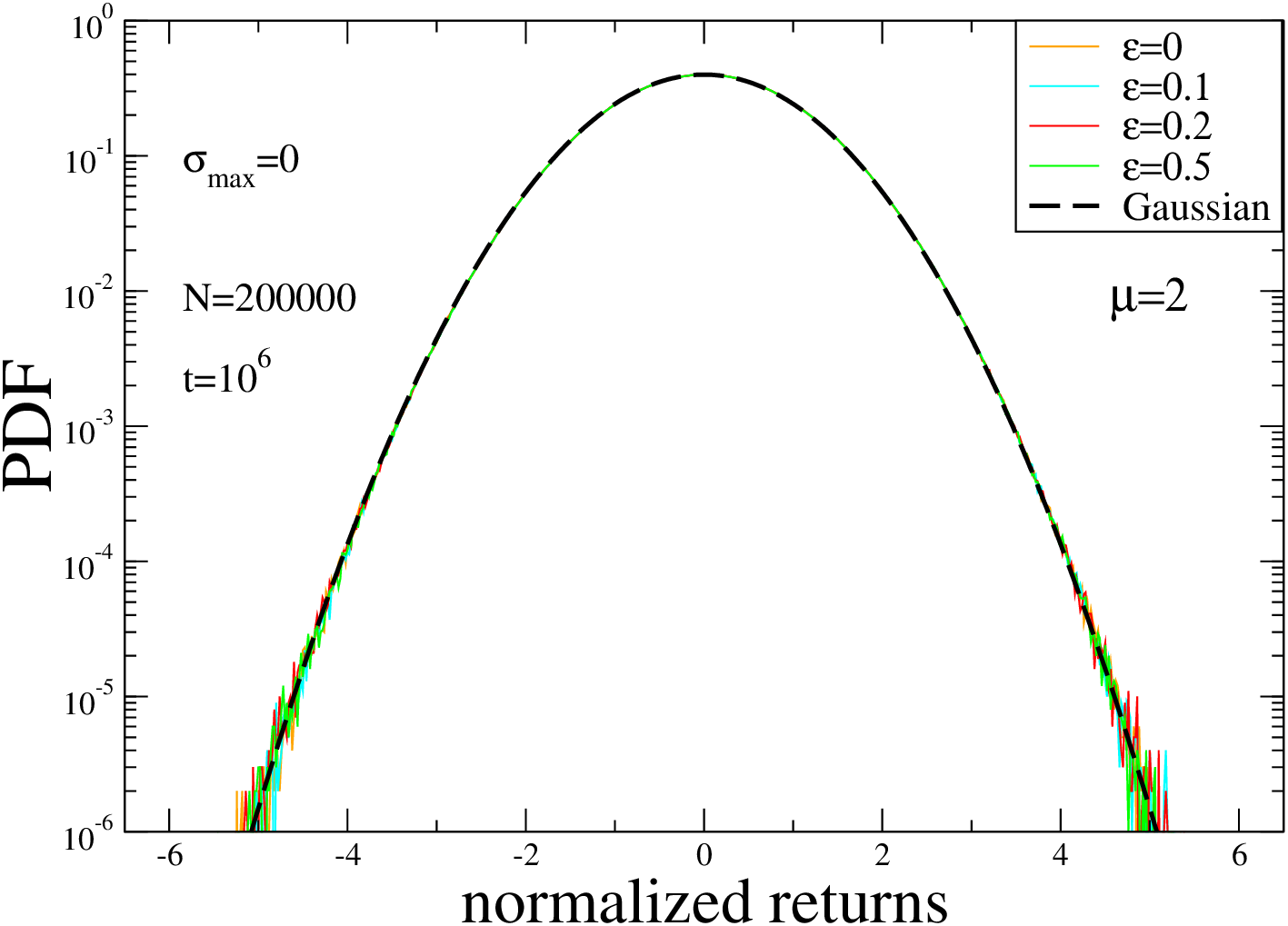}
\caption{The behaviour of the PDF for a typical case of $\mu=2$.}
\label{fig:fig9}
\end{center}
\end{figure}

\section{Conclusions}
We have numerically studied, along many decades, the time-averaged returns in the model recently 
introduced in \cite{Alex2013} by varying the number $N$ of logistic maps in the chain, 
the first-neigbour-coupling constant $\epsilon$, the width $\sigma_{max}$ of the common noise, 
and the characteristic return time $\tau$.

Our results can be summarised as follows. When the control parameter $\mu$ of the logistic maps is taken 
to be close to the edge of chaos ($\mu_c$), where the Lyapunov exponent {\it vanishes}, 
slightly asymmetric distributions are observed for some parameter values. 
Unless the number of decades is quite large, this asymmetry can be neglected and the distributions 
satisfactorily admit $q$-Gaussian approximants even for these parameter values. 
We have illustrated the effect on the index $q$ of the various parameters of the model.

When we choose for the parameter $\mu$ of the logistic maps not its value at the edge of chaos but 
any other value such that the Lyapunov exponent is {\it positive}, we obtain numerical PDFs that, 
as expected, are well fitted by simple Gaussians along six PDF decades when the system 
has no noise. In the presense of noise, however, the system is unable to achieve the Gaussian 
PDF due to the persistent perturbation of the noise. 

The overall scenario is that, when we consider time-averaged relevant quantities such as the returns, 
Boltzmann-Gibbs statistics (with its Gaussians, consistent with the classical Central Limit Theorem) 
emerges when strong chaos (positive Lyapunov exponent) is present, 
and nonextensive statistical mechanics (with its $q$-Gaussians, consistent with the $q$-generalized 
Central Limit Theorem \cite{qCLT}) emerges, or nearly emerges, when weak chaos (zero Lyapunov exponent) 
is present. These facts reinforce the results recently obtained with time-averaged quantities for 
the (conservative) standard map \cite{TirnakliBorges2014}. Biological and other systems appear to be 
well mimicked by the Pluchino et al model \cite{Alex2013}. The present results might be useful for 
examining such systems.

\section*{Acknowledgment}
This work has been supported by TUBITAK (Turkish Agency) under the Research Project number 112T083. 
U.T. is a member of the Science Academy, Istanbul, Turkey. One of us (CT) acknowledges partial support 
from CNPq, Capes and Faperj (Brazilian agencies) and the John Templeton Foundation.


\end{document}